\documentclass{pasa}%
\usepackage{graphicx}
\usepackage{gensymb}
\usepackage{multirow}

\usepackage{caption}
\usepackage{subcaption}
\usepackage[flushleft]{threeparttable}

\usepackage{aas_macros}
\usepackage{hyperref}
\hypersetup{colorlinks,citecolor=blue,linkcolor=blue,urlcolor=blue}

\newcommand{\correct}[1]{{#1}}

\newcommand{\Ofour}{O4}

\title{The capability of the Australian Square Kilometre Array Pathfinder to detect prompt radio bursts from neutron star mergers}

\author[Wang et al.]{Ziteng Wang$^{1,2,3}$\thanks{ziteng.wang@sydney.edu.au}, Tara Murphy$^{1,3}$, David L.\ Kaplan$^4$, Keith W. Bannister$^{2}$ and Dougal Dobie$^{1,2,3}$
$^1$Sydney Institute for Astronomy, School of Physics, University of Sydney, Sydney, New South Wales 2006, Australia.

$^2$CSIRO Astronomy and Space Science, PO Box 76, Epping, New South Wales 1710, Australia

$^3$ARC Centre of Excellence for Gravitational Wave Discovery (OzGrav), Hawthorn, Victoria, Australia

$^4$Center for Gravitation, Cosmology, and Astrophysics, Department of Physics, University of Wisconsin-Milwaukee, P.O. Box 413, Milwaukee, WI 53201, USA
}

\jid{PASA}
\doi{10.1017/pas.\the\year.xxx}
\jyear{\the\year}

\begin{document}

\begin{frontmatter}
\maketitle

\begin{abstract}
We discuss observational strategies to detect prompt bursts associated with gravitational wave events using the Australian Square Kilometre Array Pathfinder (ASKAP).
Many theoretical models of binary neutron stars mergers predict that bright, prompt radio emission would accompany the merger.  The detection of such prompt emission would greatly improve our knowledge of the physical conditions, environment, and location of the merger.  However, searches for prompt emission are complicated by the relatively poor localisation for gravitational wave events, with the 90\% credible region reaching hundreds or even thousands of square degrees.
Operating in fly's eye mode, the ASKAP field of view can reach $\sim$1000 deg$^2$ at $\sim 888\,{\rm MHz}$. This potentially allows observers to cover most of the 90\% credible region quickly enough to detect prompt emission.
We use  skymaps for GW170817 and GW190814 from LIGO/Virgo's third observing run to simulate the probability of detecting prompt emission for gravitational wave events in the upcoming fourth observing run.   With only alerts released after merger we find it difficult to slew the telescope sufficiently quickly as to capture any prompt emission.  However, with the addition of alerts released \textit{before} merger by negative-latency pipelines we find that it should be possible to search for nearby, bright prompt FRB-like emission from gravitational wave events.  Nonetheless, the rates are low:  we would expect to observe $\sim$0.012 events during the fourth observing run, \correct{assuming that the prompt emission is emitted microseconds around the merger}.
\end{abstract}

\end{frontmatter}

\section{INTRODUCTION }
\label{sec:intro}

In 2017, the Advanced Laser Interferometer Gravitational-Wave Observatory  and Virgo Interferometer \citep[aLIGO/Virgo;][]{Aasi_2015,AdV} detected gravitational waves (GWs) from a binary-neutron-star (BNS) merger, GW170817 \citep{PhysRevLett.119.161101}, followed by the detection of gamma rays 1.7 seconds later \citep[GRB 170817A;][]{Abbott_2017a}. It was the first joint detection of gravitational waves and electromagnetic radiation from the same source. However, the delay in issuing the alert and large error region prevented most  telescopes from searching for any prompt transient event \correct{microseconds around the merger}. 

Many models \correct{(including some that predate the discovery of fast radio bursts \citep[FRBs;][] {Lorimer777,Thornton53,2019ARA&A..57..417C})} predict prompt radio emission associated with compact object mergers. This emission could be generated by the magnetic field interactions during the inspiral \citep{Hansen2001,Lai_2012,Totani_2013,Metzger_2016,Wang_2016,Wang_2018},
interaction between a relativistic jet and interstellar medium \citep{Pshirkov_2010},
or the collapse of a supramassive neutron-star remnant into a black hole \citep{Ravi_2014,Falcke2014}.
In particular, prompt emission may be in the form of short coherent radio pulses like FRBs. While some estimates suggested that neutron star mergers could not be the sole progenitors of FRBs because the volumetric rate of FRBs is significantly higher than that of BNS mergers, $\gtrsim 10^4\,{\rm Gpc}^{-3}\,{\rm yr}^{-1}$ \citep{Ravi2019,2018ApJ...858...89C} versus $1540_{-1220}^{+3200}\,{\rm Gpc}^{-3}\,{\rm yr}^{-1}$ \citep[][also see \citealt{2016ApJ...825L..12C}]{Abbott_2017a}, other estimates of the FRB rate have found better agreement \citep{2019ApJ...883...40L}. 
\correct{However, the detection of repeating FRBs \citep[e.g.:FRB121102][]{FRB121102} clearly demonstrates that not all FRBs originate from BNS mergers.}
Further complications have come from recent observations, when the Galactic magnetar SGR 1935+2154 had X-ray outbursts \citep{2020GCN.27668....1M,2020arXiv200506335M,2020GCN.27669....1R,2020arXiv200511178R,2020ATel13687....1Z} accompanied by simultaneous radio bursts \citep{2020ATel13681....1S} of a brightness consistent with faint extragalactic FRBs \citep{2020arXiv200510828B,2020arXiv200505283M}.  
\correct{\cite{james_2019}, \cite{2018MNRAS.474.1900M}, and \cite{2019MNRAS.483.1342J} also note that the FRB distribution we have got so far may be different by taking statistical and systemic effects into account.} 
As is stands, the fraction of non-repeating FRBs potentially caused by BNS mergers is unclear, but a single discovery would change that.
Furthermore, the detection of prompt emission from a binary neutron star merger would be a useful tool to measure the interstellar medium and  magnetic environment \citep{2019MNRAS.483..359L} near the merger and locate the progenitor quickly, going from the 10--1000\,deg$^2$ localisation from GW alone \citep{Singer_2014,GW190814} to $\sim$arcmin \citep{Bannister_2017} for FRBs localised by the Australian Square Kilometre Array Pathfinder \citep[ASKAP;][]{ASKAP}. Combining this localisation with distance constraints from GW pipelines may be sufficient to uniquely identify a host galaxy, even in the absence of sub-arcsecond localisation.

The timing of radio emission relative to the merger varies between different models. \cite{Totani_2013} predicts that FRBs produced by the magnetic field interactions could occur milliseconds before the BNS merger, while \cite{Falcke2014} predict an FRB due to the collapse of the supramassive neutron star produced 10 to 10\,000 seconds after the merger. Hence to detect or put limits on this prompt emission, we need to cover the GW error region as quickly as possible,  ideally before the merger occurs, and then continue observing for minutes to hours after.

In late 2021, a four-detector GW network with the two aLIGO instruments, aLIGO Hanford (H) and aLIGO Livingston (L) combined with Phase 1 of Advanced Virgo (V)  and the Kamioka Gravitational Wave Detector \citep[KAGRA, or K;][]{KAGRA_1,KAGRA_2} will be online for the fourth GW observing run \citep[O4;][]{Abbott2018}, with BNS detectability ranges from 160--190\,Mpc (L, H) to 90--120\,Mpc (V) to 25--130\,Mpc (K).  During O4 the network is expected to detect between 3 and 110 BNS mergers using the estimated BNS event rate of 110--3840\,Gpc$^{-3}$\,yr$^{-1}$ scaled from the observation of GW170817 \citep{Abbott_2018b}.

In contrast to O3 and earlier, where GW alerts were released only after the mergers with latencies of minutes to hours, 
in the O4 observing period it is expected that at least one negative-latency pipeline will be operating on the LHVK network. Negative-latency pipelines use matched filters to detect GW signals in the inspiral phase, which enables an alert to be released before the BNS merger occurs \citep{Cannon_2012,Chu_2016}. \cite{james_2019} illustrated that summed parallel infinite impulse response \citep[\texttt{SPIIR};][]{SPIIR1,SPIIR2} can detect GW170817-like events about 30 seconds before the merger with detection signal to noise ratio (SNR) of $\sim$10.
\cite{Chu_2016} found that the median localization error areas can be better than 500 deg$^2$ with a four-detector network 40 seconds before the coalescence.

There have been several unsuccessful attempts to search for prompt radio emission from compact binary coalescence at low radio frequencies. \cite{Callister_2019} used the Owens Valley Radio Observatory Long Wavelength Array (OVRO-LWA, observing at 27--85\,MHz)\footnote{\url{http://www.tauceti.caltech.edu/LWA/}}
 to search for the prompt emission from the binary black hole merger GW170104, while \citet{2018ApJ...864...22A} used the OVRO-LWA to search for prompt emission from the cosmological short gamma-ray burst 170112A.  Similarly, \citet{2019MNRAS.489.3316R} examined data \citep{2015ApJ...814L..25K} from the Murchison Widefield Array \citep[MWA;][observing at 70--300\,MHz]{MWA} from the  short gamma-ray burst 150424A.  
 \correct{However, these events are well beyond the BNS detection sensitivity distance in \Ofour ($z\sim 0.05$)}.
 More broadly,
\cite{kaplan_2016} investigated  strategies  to observe prompt emission from GW events with the MWA (also see \citealt{james_2019}).

\begin{figure}
    \centering
    \includegraphics[width=\columnwidth]{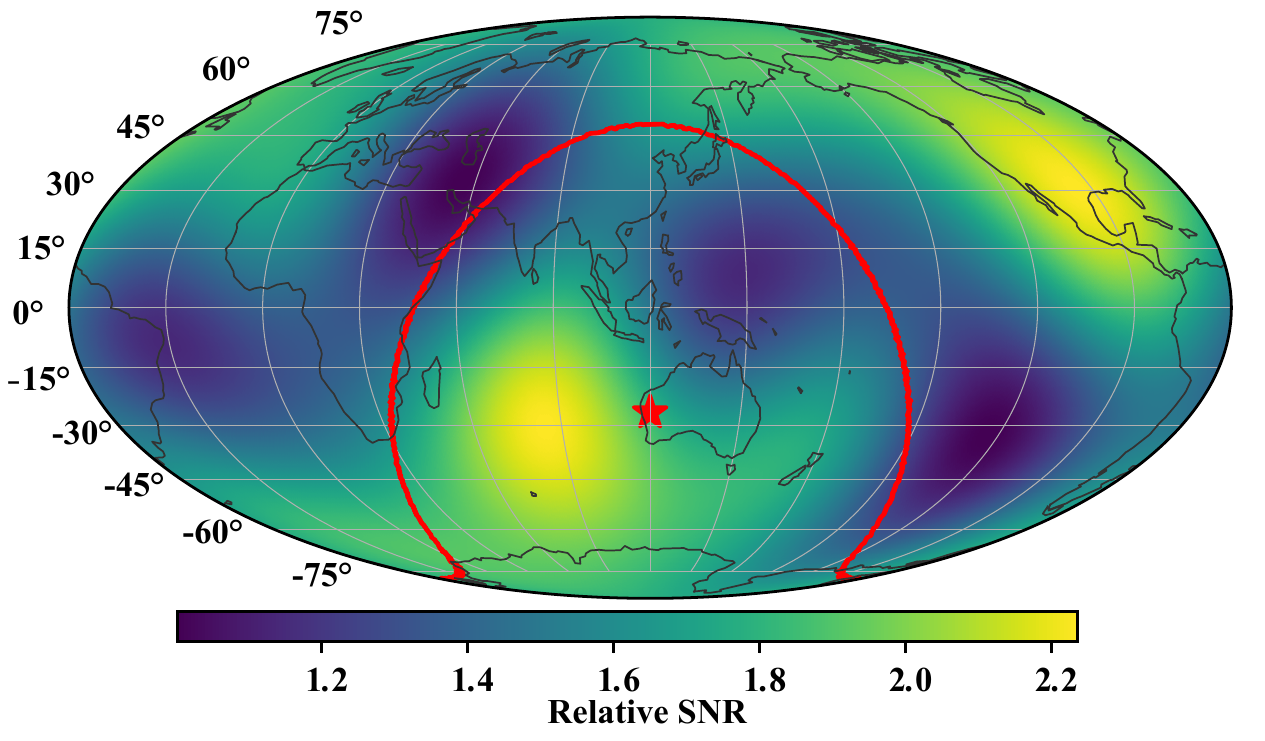}
    \caption{Sensitivity map of the gravitational wave detector network in \Ofour (HLVK). The color is proportional to the relative signal-to-noise ratio. The red line shows the ASKAP horizon (15$^\circ$ in elevation angle) and the red star is where ASKAP is located.}
    \label{fig:O4_sens}
\end{figure}

The Australian Square Kilometer Array Pathfinder \citep[ASKAP;][]{ASKAP} is a $36 \times 12$-m antenna radio telescope located in Western Australia, operating over the frequency range 700\,MHz to 1.8\,GHz.
In 2019 we conducted follow-up observations of several GW events with ASKAP in its standard imaging mode \citep[e.g., ][]{Dobie_2019b}.  In parallel, 
the Commensal Real-time ASKAP Fast Transients survey \citep[CRAFT;][]{CRAFT} has developed high time resolution
(1\,ms) capabilities on ASKAP, which has resulted in the discovery of at least 32 FRBs \citep[e.g.,][]{Shannon2018, Bannister_2017, Bannister565}. 
The localisation precision for ASKAP can be $\sim$arcminute with probabilistic localisation techniques \citep[e.g.,][]{Bannister_2017}, going to $\sim$arcsecond using standard interferometry techniques \citep[e.g.,][]{Bannister565}. 
\correct{ASKAP covers sky regions of a higher detection sensitivity for the gravitational wave detector network than some other radio telescopes (we show the corresponding sensitivity map in Fig. \ref{fig:O4_sens}). This enables ASKAP to detect gravitational wave signals earlier. However, the differences will be smaller in the future with more detectors joining the network.}

In this paper we discuss the capabilities of ASKAP for detecting prompt emission from gravitational wave events.

\begin{figure*}[h]
\begin{subfigure}{.5\textwidth}
  \centering
  \includegraphics[width=\columnwidth]{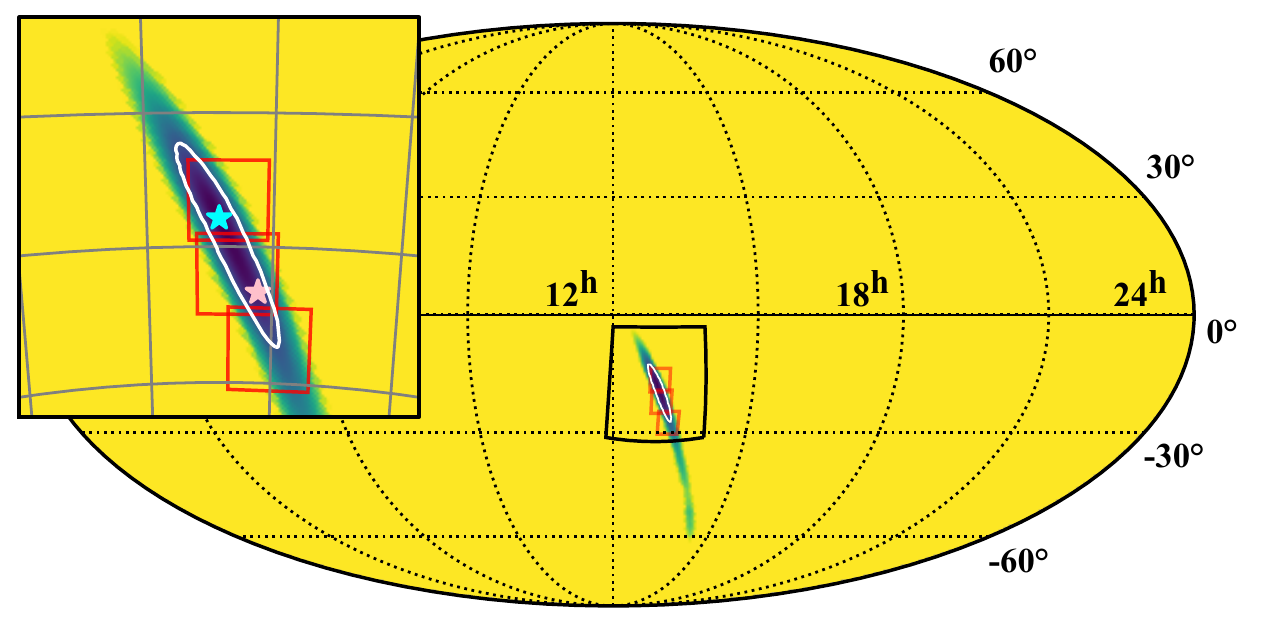}
\end{subfigure}%
\begin{subfigure}{.5\textwidth}
  \centering
  \includegraphics[width=\columnwidth]{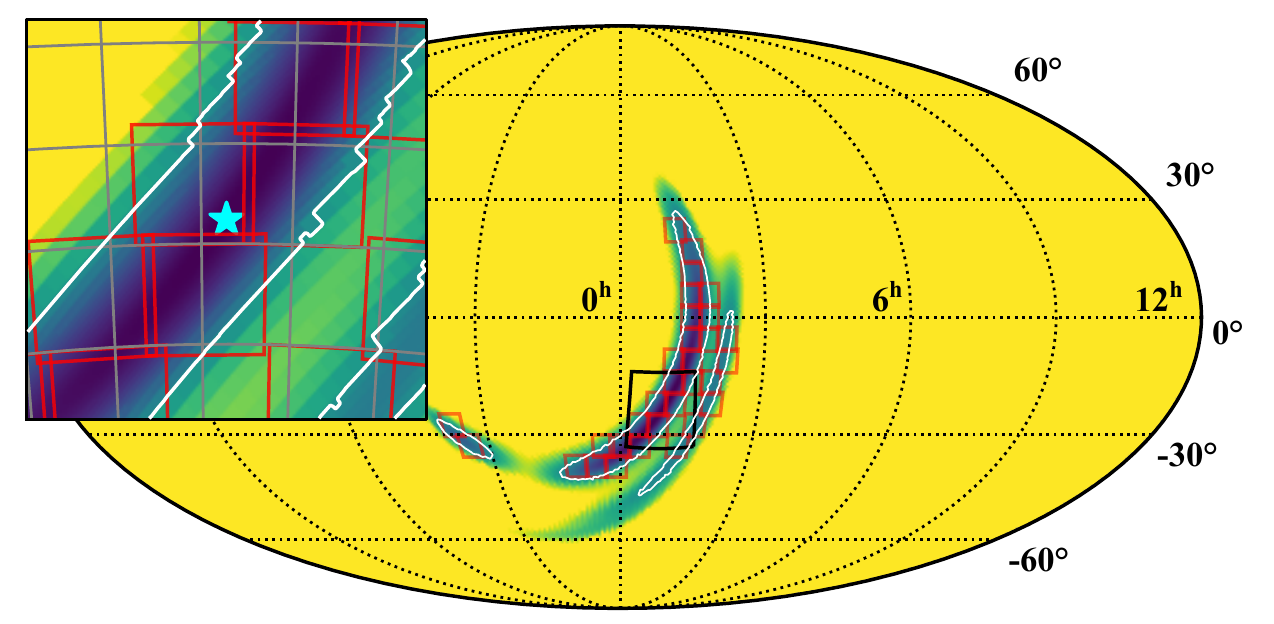}
\end{subfigure}
\caption{Sky localisation maps for GW170817 (\protect\citealt{Abbott_2017a}; left) and the initial map for GW190814\protect\footnotemark (\protect\citealt{GW190814}; right). The color is proportional to the log$_{10}$ of the probability.  Maps are plotted in equatorial coordinates using the Mollweide projection.  
The 90\% credible regions are shown by the white lines.
The optimised tilings of ASKAP \protect\citep{dobie_2019} to cover 90\% credible region are shown in the red squares. Each square is 6$\times$6\,deg$^2$ FoV of ASKAP.
The number of ASKAP tilings for GW170817 is 3 while that for GW190814 is 35. 
The cyan star shows the position with the maximum posterior probability for both events and the pink star shows where GW170817 actually was.
Zoomed regions show the insets around the maximum posterior probability positions. The sizes for zoomed regions are 30$\times$30\,deg$^2$ (left) and 20$\times$20\,deg$^2$ (right) respectively.}
\label{fig:GW170817}
\end{figure*}
 
\footnotetext{\url{https://gracedb.ligo.org/apiweb/superevents/S190814bv/files/bayestar.fits.gz}}
 
\section{Negative-Latency Pipelines}

There are a number of pipelines running on aLIGO/Virgo that are designed to rapidly identify compact binary merger events. The key pipelines are \texttt{GstLAL} \citep{GstLAL_2017a,GstLAL_2019b}, multi-band template analysis \citep[\texttt{MBTA};][]{MBTA_2016}, \texttt{PyCBC Live} \citep{PyCBC_2017,PyCBC_2018} and \texttt{SPIIR} \citep{SPIIR1,SPIIR2}.

Each pipeline sends an alert to the \texttt{}{Gravitational-Wave Candidate Event Database} (GraceDB)\footnote{\url{https://gracedb.ligo.org/}} if the detection statistic (e.g., signal-to-noise ratio) of the candidate passes a predetermined threshold.
In order to collect the information about a single physical event in one place and issue a single alert per physical event, GraceDB gathers individual alerts into a ''superevent'' data model.\protect\footnote{See \url{https://gracedb.ligo.org/documentation/models.html\#superevents} and \url{https://emfollow.docs.ligo.org/userguide/analysis/superevents.html}}
Pipelines which send out superevents quickly will facilitate searches for prompt emission from BNSs.

Bayesian triangulation and rapid localization \citep[\texttt{BAYESTAR};][]{Singer_bayestar} algorithms can offer observers a rapid parameter estimation for the GW events including the event localisation (e.g., Figure~\ref{fig:GW170817} for GW170817), distance and component masses estimates.  In general, increasing from a two-detector network to a three-detector network (O3) to a  four-detector network operating in O4 will improve localisation \citep[e.g.:][]{Chu_2016,Abbott_2017a}.

However, moving to a negative latency search will reduce the SNR at detection and hence increase the location uncertainty. 
In general, the earlier the detection is, the lower SNR the pipeline will determine and the worse the localisation the pipeline will produce \citep{Chu_2016}.  \cite{james_2019} shows that the SNR of the GW170817 signal at $t_0- 30\,$ (i.e., 30\,s before the merger at $t_0$) is one third of that at the time of the merger. \cite{Cannon_2012} and \cite{Chu_2016} show that the localization can be thousands of square degrees at $t_0-30\,$s,  even if it  is less than 30\,deg$^2$ at $t_0$.
\cite{Chu_2016} used simulated data to calculate the median 90\% credible regions at different times prior to merger for different detector networks.
The 90\% credible region will improve from about 2200\,deg$^2$ to 1600\,deg$^2$ at $t_0-40\,$s  from O3 with the LHV network to O4 with the LHVK network.


\section{ASKAP capabilities}\label{sec:ASKAP}

\begin{table*}[h]
\caption{ASKAP Antenna Mount Operating Characteristics$^a$}
\centering
\begin{threeparttable}

\begin{tabular}{@{}ccccc@{}}
\hline\hline
Axis & Range of Motion & Rotation Range & Slew Rate & Acceleration/Deceleration Speed \\
\hline%
 Azimuth  & Full & +/$-$270 deg & 3 deg sec$^{-1}$ & 3 deg sec$^{-2}$\\
 Elevation  & +15 deg to +89 deg & N/A & 1 deg sec$^{-1}$ & 1 deg sec$^{-2}$\\ 
\hline\hline
\end{tabular}
\begin{tablenotes}
\begin{footnotesize}
\item[a] From Australia Telescope National Facility (ATNF):  \url{https://www.atnf.csiro.au/projects/askap/ASKAP_Antenna_public_specification_Nov08_v0.0.pdf}
\end{footnotesize}
\end{tablenotes}
\end{threeparttable}
\label{askap_char}
\end{table*}


ASKAP consists of 36  antennas which can point along their altitude  and azimuth axes separately, with operating characteristics in Table~\ref{askap_char}. 
The telescope can slew at a rate of 3\,deg\,sec$^{-1}$ and 1\,deg\,sec$^{-1}$ in azimuth and altitude respectively.
The rotation range for the azimuth axis is from $-270\degree$ to $270\degree$ to allow for cable wraps.
When reaching one of these limits, the antennas may need to unwrap by $\pm360\degree$  even if the target position is very close to the starting position, adding an additional overhead of $\sim$120\,s. ASKAP has an additional rotation axis where the dish rotates around the optical axis. In a typical fast-slew strategy we would choose not to change the dish rotation in order to minimise the total settling time.

\begin{table}[h]
\caption{Sensitivity, field-of-view (FoV) and angular resolution  for different ASKAP of sub-arrays}
\begin{threeparttable}
\centering
\setlength\tabcolsep{2pt}
\begin{tabular}{@{}cccc@{}}
\hline\hline
Configuration & FoV$^a$ & $\sigma_{\rm det}^b$ & Localisation \\
 & (deg$^2$) & (Jy) & Capability \\
\hline%
1-antenna sub-array & $\sim1080$ & 2.4 & {$10^\prime\times10^\prime$} \\
36-antenna sub-array & $\sim30$ & 0.4 & {$3^{\prime\prime}\times3^{\prime\prime}$} \\
\hline\hline
\end{tabular}
\begin{tablenotes}
\begin{footnotesize}
\item[a] This assumes that we can use all 36 antennas and all antennas operate in the same configuration but different sub-array points to different position.\\
\item[b] With integration time of 1\,ms, bandwidth of 336\,MHz.
\end{footnotesize}
\end{tablenotes}
\end{threeparttable}
\label{sub_array_tab}
\end{table}
There are two operational modes for ASKAP to look for FRB-like emission\footnote{As yet ASKAP cannot do a fully coherent FRB search using all antennas.}: collimated incoherent mode and fly's eye mode \citep[][and see Table~\ref{sub_array_tab}]{Bannister_2017}. 
In the collimated incoherent mode, all the antennas point in the same direction, and the total power detected at each antenna is combined incoherently.
In the fly's eye mode, each antenna points in a different
direction and we can achieve a total FoV N$\times$30\,deg$^2$ with N antennas, albeit with reduced sensitivity.   

There is also a mode between the collimated incoherent mode and fly's eye mode where one can use the 36 antennas in  separate incoherent sub-arrays.
For example, 36 antennas can be operated simultaneously as six separate 6-antenna arrays, with each sub-array producing a separate data stream.
Fly's eye mode is equivalent to 1-antenna sub-arrays, while full collimated incoherent mode is corresponding to 36-antenna sub-array.

The detection uncertainty, $\sigma_{\rm det}$, for a single ASKAP antenna is
\begin{equation}
    \sigma_{\rm det} = \frac{\textrm{SEFD}}{\sqrt{\Delta \nu t N_{\rm pol}}},
\label{ant_sig}
\end{equation}
\\where SEFD is the system equivalent flux density ($\sim$2000\,Jy for ASKAP)\footnote{\url{https://www.atnf.csiro.au/projects/askap/memo015_a.pdf}}, $\Delta\nu=336\,$MHz is the bandwidth, $t$ is the integration time and $N_{\rm pol}=2$ is the number of the polarisations. Typical FRB searches use $t=1\,$ms. Using the equation above, the noise for a single ASKAP  dish for an FRB search is roughly 2.4\,Jy. When combining multiple antennas incoherently the noise decreases as $1/\sqrt{N_{\rm ant}}$, with $N_{\rm ant}$ the number of antennas.
Therefore the noise can reach $\sim 0.4$\,Jy for the collimated incoherent mode when using all 36 antennas.

\section{Slewing Simulations}\label{sec:sim}

Most models predict that prompt emission (if present) would occur within a few seconds of the merger \citep[e.g.,][]{Hansen2001}.
The chance of detecting this prompt emission will depend on both the latency of the pipelines and the slew time of the  antennas. If the latency plus the slew time is smaller than the delay before radio emission, it may be possible to detect prompt emission.

Given initial and final positions of each antenna, we can use Table~\ref{askap_char} to calculate the slew time for both axes (including acceleration and deceleration) and take the maximum of the two axis times as the slew time of each antenna. However, since we consider observation modes where all of the antennas end up pointing in different positions, we take the maximum of the slew times of all antennas as the slew time for the event.

We test this with two limiting cases: (1) a small event region where all of the antennas point to the same location (credible region $\lesssim$ ASKAP field of view); (2) a large event region where we use the fly's eye mode to tile as much as we can.  Since the starting positions are unknown, we simulate this with randomized starting positions but two concrete sky maps: GW170817 \citep{Abbott_2017a} as a small region ($\sim 30\,{\rm deg}^2$), and GW190814 as a large region ($\sim 772\,{\rm deg}^2$), as illustrated in Figure~\ref{fig:GW170817}.
We use the initial {\sc bayestar} skymap for GW190814 \citep{GCN25324} as an example of a poor early-time localisation (although we note that the final \texttt{LALInference} skymap had a smaller 90\% localisation of  $\sim30\,{\rm deg}^2$ which was similar to that for GW170817). The 90\% localisation in Figure~\ref{fig:GW170817} is significantly larger than the ASKAP FoV and is elongated (like most GW localisations) and therefore it  serves as a useful illustration.

The starting conditions were randomised with equal probability for all azimuths and uniform in cosine of the elevation angle (equal probability per sky area) between $15\degree$ and $89\degree$.

\begin{figure*}[!h]
  \centering
  \includegraphics[width=\textwidth]{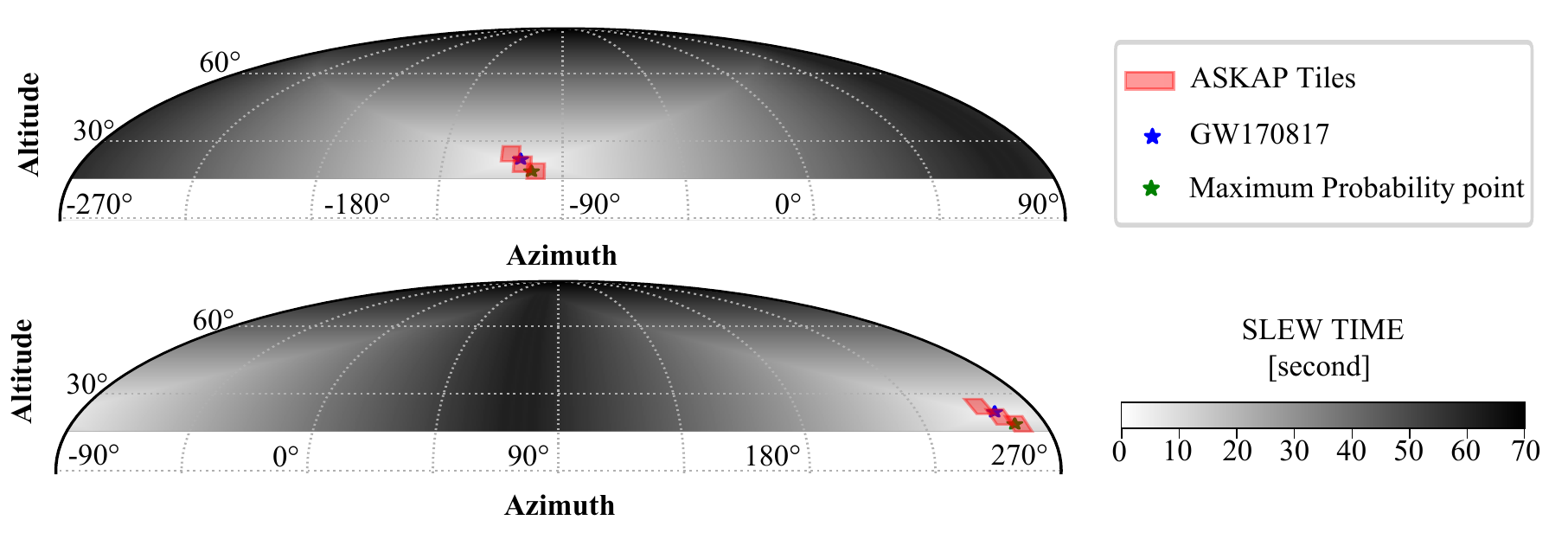}
\caption{Slew time (indicated by the grey scale) for different starting position for GW170817 as a function of starting azimuth and altitude. The red rectangles show the tiling for the \texttt{bayestar} skymap from the HLV network. The blue star shows the point where GW170817 is and the green star shows the maximum probability position of the skymap. The figure is in the azimuth/altitude coordinates and is modified to cover azimuths from $-270$\,degree to $270$\,degree, allowing for antenna motion over that range.}

\label{fig:GW170817sim}
\end{figure*}

\begin{figure}
    \centering
    \includegraphics[width=\columnwidth]{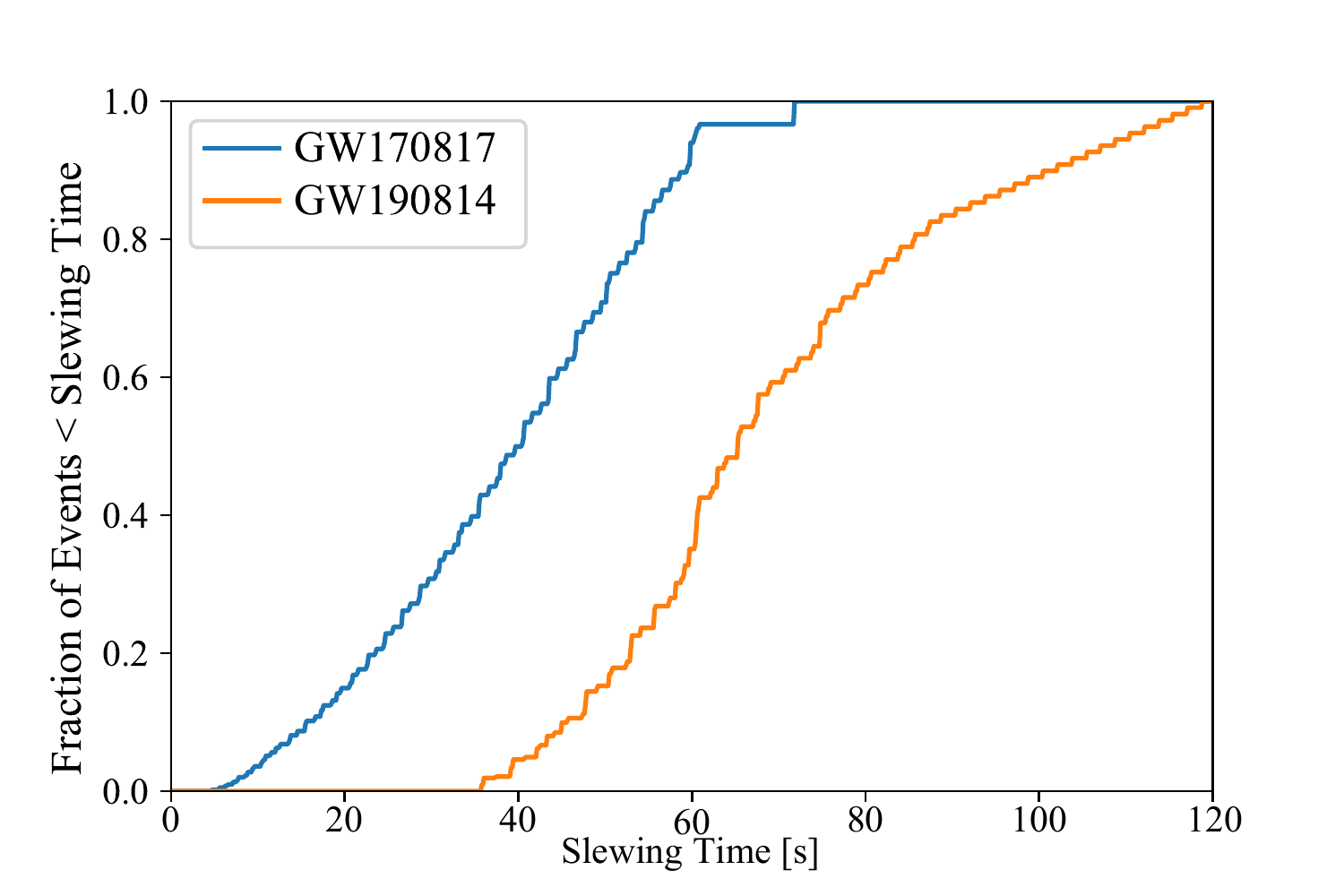}
    \caption{Cumulative histogram of slew times for GW170817 (blue) and GW190814 (orange) for different initial positions. The y-axis shows the fraction of the events whose slew time is smaller than the value on the x-axis.}
    \label{fig:slewingSim}
\end{figure}

For the small region, GW170817 ($\sim 30\,{\rm deg}^2$), we can cover the 90\% credible region with three FoVs (shown in Figure~\ref{fig:GW170817} and Figure~\ref{fig:GW170817sim}). Figure~\ref{fig:GW170817sim} shows the spatial distribution of slewing time for GW170817. The median slewing time to get on-source is 40.3\,s, with a range of 15.5--59.3\,s for the 10th and 90th percentiles, respectively,  as shown in Figure~\ref{fig:slewingSim}.


For the large region, we used the strategy in \cite{dobie_2019} to optimise the ASKAP pointing. 
For simplicity, we called the coverage of a single antenna tiling.
This strategy optimises the tiling center position to tile as much of the  90\% credible region as ASKAP can. 
If more than the maximum 36 antennas are needed to tile the 90\% credible region, then less area will be covered.  However, if we can cover the 90\% region with fewer than 36 antennas then more than one antenna is pointed toward the highest probability region.  
We used the GW190814 {\sc bayestar} localisation to perform our simulation for the large region. We need all 36 antennas to cover the 90\% credible region. As is shown in Figure~\ref{fig:slewingSim}, the 50th and 90th percentile slewing times to get on-source are 65.2\,s and 102.0\,s.

\section{Discussion}

There could  be as many as tens of BNS mergers detected by the HLVK gravitational wave detector network in the O4 run   \citep{Abbott2018}. It may be impossible to follow up every event, especially including false alarms that increased during the low-latency portions of O3 and might increase further with negative latencies.

Only high signal-to-noise GW events will be amenable to negative latency detections, such that the reduced signal-to-noise ratio will still pass the detection threshold.  Overall, higher signal-to-noise ratio detections lead to smaller localisations, with $A \propto \left(\rho \sigma_f\right)^{-2}$ where $\rho$ is the signal-to-noise ratio and $\sigma_f$ is the effective bandwidth of the source in the detector \citep{Fairhurst_2009}.  While high-$\rho$ events will have well-localised detections eventually,  moving to negative latency will reduce the instantaneous $\rho$ and $\sigma_f$, and therefore increase the sky area accordingly. 
At the threshold where $\rho\geq 8$ at $t_0-30\,$s, the first localisation map would have an area of $\sim \mathrm{few}\times 10^3\,{\rm deg}^2$, decreasing to a few ${\rm deg}^2$ for the final detection with $\rho\sim 25$ \citep{Cannon_2012}.

We now examine how a likely scenario could play out for a prompt ASKAP search. 
The typical time for downloading the skymap and calculating the pointings is of the order of 1\,s.
We examine a nominal event occuring at a distance of $\sim 130\,$Mpc, with an early-warning $\rho=8$ at $t_0-30\,$s and a 90\% localisation of $\sim 5000\,{\rm deg}^2$.  This would imply a slew time of $60-90$\,s, similar to the GW190814 case in Section~\ref{sec:sim}. The dispersion measure, induced by the intergalactic medium, of events at this distance is $363\,{\rm pc}\,{\rm cm}^{-3}$ \citep{james_2019} and therefore a time-delay to an ASKAP frequency of 900\,MHz of only 1.9\,s.
\correct{The time-delay due to the intergalactic medium is relatively small compared to the slewing time, and in this case the time-delay can be considered negligible} 
ASKAP will therefore miss this event by $\gtrsim 1$\,minute. 
However, we note that based on Figure~\ref{fig:slewingSim} there are about 10\% of events where the slew time will be $\lesssim 50$\,s, in which case we may be able to conduct a search for pre-merger prompt emission.

An ideal event for our search would have an early warning sky area which is only a single ASKAP FOV (comparable to our simulation of the  GW170817 skymap in Figures~\ref{fig:GW170817} and \ref{fig:slewingSim}).  Such an event would have a final $\rho\gtrsim 60$ (at the time of the merger), corresponding to sources closer than $\sim 24\,{\rm Mpc}$ \citep{Cannon_2012}, or  a rate of $0.1^{+0.2}_{-0.1}\,{\rm yr}^{-1}$  based on the BNS rate in \cite{Abbott_2017a}.  Therefore, the prospects of seeing such an event during a year-long observing campaign are $\sim 0.1$.  For any realistic scenario we will of course lose 60\% of events to those detected below the ASKAP horizon, as well as $\sim 30$\% (assuming the alert for the event with small early-warning skymap is released 30\,seconds before the merger) of the remaining events to those where the slew-time is unacceptably long, meaning that we may only trigger on 12\% of these very bright events. So we expect that ASKAP may be able to observe $\sim 0.012$ events during O4.
At least at radio wavelengths we are not sensitive to weather and significantly less sensitive to day/night distinctions.

The discussion above focused on whether ASKAP could \textit{observe} prompt emission, but did not consider whether or not it could \textit{detect} it.  For that we need to understand the flux density predictions of different models.
Some models \citep[e.g.,][]{Pshirkov_2010,Hansen2001} predict that BNS mergers can produce FRB-like emission at low radio frequencies, $\lesssim 100\,$MHz, but this is too low for ASKAP, and some of them have a hard cutoff at the high frequency end \citep[e.g.:][]{Wang_2018}.
Moreover, no FRBs have been detected at those frequencies \citep[e.g.,][although new detections are coming closer, \citealt{2020ApJ...896L..40P,2020arXiv200402862C}]{2016MNRAS.458.3506R,2017ApJ...844..140C,2015AJ....150..199T,2018ApJ...867L..12S}, which may influence the BNS/FRB rate comparison \citep[e.g.,][]{2016ApJ...825L..12C}.
\cite{Wang_2016}, \cite{Ravi_2014} and \cite{Lyutikov_2013} predict the total power that can be emitted by BNS mergers. However, the flux density at a given frequency is not clear.  Instead we parameterise this as $4\pi d^2 \nu S_\nu = x L$, where $x$ depends on the spectrum of the emission and the upper and lower frequency limits.
In \cite{Totani_2013}, the signal is predicted to be $\sim 10^3\,{\rm Jy}$ at $24\,{\rm Mpc}$, at a frequency of $888\,{\rm MHz}$ (assuming a radio emission efficiency of $\epsilon\sim 10^{-3}$, magnetic field strength of $B\sim 10^{12.5}\,{\rm G}$,   neutron star radius of $R\sim 10\,{\rm km}$, and rotation periods of $P\sim 0.5\,{\rm ms}$).
According to Table~\ref{sub_array_tab}, if  the merger follows the mechanism shown in \cite{Totani_2013}, even with fly's eye mode ASKAP can detect the signal if covering the burst region. 
As mentioned above, BNSs within 24\,Mpc are estimated to occur infrequently, at a rate of $0.1^{+0.2}_{-0.1}\,{\rm yr}^{-1}$. Though the expected flux density is high, the rate for such a bright event is low, which makes it harder to detect any prompt emission.
For the other models we could detect them if $x>3\times10^{-6}$ (assuming  a radio luminosity of $L\sim10^{40}\,{\rm erg}\,{\rm s}^{-1}$ as in \citealt{Wang_2016}, an event distance of $d\sim24\,{\rm Mpc}$, and on observing frequency of 888\,MHz).


\correct{In this paper, we have mainly focused on BNS mergers, although one of our sky maps in Figure~\ref{fig:GW170817} was actually from a neutron-star black-hole  (NSBH) merger.  We generally consider BNS mergers as more favorable to production of EM counterparts in general and FRB-like emission in particular, 
although \cite{Mingarelli_2015} hypothesizes that NSBH mergers  could also produce FRBs. However, NSBH mergers will generally contain only a single magnetosphere and in many cases the NS will be swallowed by the BH without disruption \citep{Mingarelli_2015,2020ApJ...896...54C}, so many models for prompt emission will fail. Separately, NSBH mergers will generally be more distant \citep[e.g.,][]{Abbott_2018b}, leading to fainter EM emission, and the poorly-constrained merger rate \citep{GW190814}
makes predictions difficult.
Regardless, ASKAP still can and will search for the prompt emission from a well-localised NSBH merger, because the basic detection scenarios in terms of localisation will be similar to that for BNS mergers. 
}

\section{Conclusion}

The detection of prompt emission from BNS mergers requires a radio instrument of sufficient sensitivity that is capable of being on-target before the arrival of a burst. 
In this paper we estimated the probability for ASKAP to capture the prompt emission with negative-latency alerts from aLIGO/Virgo using two observational modes suitable for FRB-like emission.  
Fly's eye mode can achieve higher sky coverage but less sensitivity, while the collimated incoherent mode gains sensitivity at the expense of sky area.
Given that faint, poorly localised events will also have minimal advance warning and so will not allow for sufficient slew times, it may be nearly impossible to capture any prompt emission from those events.   However, brighter events will have both better advance warning and smaller localisations areas enabling more sensitive observations and more confident coverage of the whole localisation area: we can achieve better sensitivity (down to 3.2\,Jy using all antennas incoherently with an 8$\sigma$ detection threshold). If FRB-like emission can be detected from these events, we can get a more accurate localisation for the event (as small as 3$^{\prime\prime}\times$3$^{\prime\prime}$ with more than 3 antennas with the same pointing), which can enable followup across the electromagnetic spectrum.
While expected to be rare ($\sim 0.012\,{\rm yr}^{-1}$), ASKAP observations of negative-latency triggers have the potential to discover prompt emission from BNS mergers or at least constrain their origin.


\begin{acknowledgements}

We thank Prof. Jolien Creighton and Dr. Deep Chatterjee for helpful discussions.
We thank the referee for his/her thorough review and highly appreciate the comments and
suggestions, which significantly contributed to improving the quality of the publication. 
TM acknowledges the support of the Australian Research Council through grant DP190100561. DD is supported by an Australian Government Research Training Program Scholarship.  DK is supported by NSF grant AST-1816492.
Parts of this research were conducted by the Australian Research Council Centre of Excellence for Gravitational Wave Discovery (OzGrav), project number CE170100004.
The Australian SKA Pathfinder is part of the Australia Telescope National Facility which is funded by the Commonwealth of Australia for operation as a National Facility managed by CSIRO. The Murchison Radio-astronomy Observatory is managed by the CSIRO, who also provide operational support to ASKAP. We acknowledge the Wajarri Yamatji people as the traditional owners of the Observatory site. This research has made use of NASA's Astrophysics Data System Bibliographic Services.

\end{acknowledgements}

\bibliographystyle{pasa-mnras}
\bibliography{Reference}

\begin{thebibliography}{}
\makeatletter
\relax
\def\mn@urlcharsother{\let\do\@makeother \do\$\do\&\do\#\do\^\do\_\do\%\do\~}
\definecolor{darkblue}{rgb}{0,0,0.597656}
\def\mndoi{\begingroup\mn@urlcharsother \@ifnextchar [ {\mndoi@} {\mndoi@[]}}
\def\mndoi@[#1]#2{\def\@tempa{#1}\ifx\@tempa\@empty \href
  {http://dx.doi.org/#2} {\textcolor{darkblue}{doi:#2}}\else \href
  {http://dx.doi.org/#2} {\textcolor{darkblue}{#1}}\fi \endgroup}
\def\mn@eprint#1#2{\mn@eprint@#1:#2::\@nil}
\def\mn@eprint@arXiv#1{\href {http://arxiv.org/abs/#1} {{\tt arXiv:#1}}}
\def\mn@eprint@dblp#1{\href {http://dblp.uni-trier.de/rec/bibtex/#1.xml}
  {dblp:#1}}
\def\mn@eprint@#1:#2:#3:#4\@nil{\def\@tempa {#1}\def\@tempb {#2}\def\@tempc
  {#3}\ifx \@tempc \@empty \let \@tempc \@tempb \let \@tempb \@tempa \fi \ifx
  \@tempb \@empty \def\@tempb {arXiv}\fi \@ifundefined
  {mn@eprint@\@tempb}{\@tempb:\@tempc}{\expandafter \expandafter \csname
  mn@eprint@\@tempb\endcsname \expandafter{\@tempc}}}

\bibitem[\protect\citeauthoryear{Aasi et~al.,}{Aasi et~al.}{2015}]{Aasi_2015}
Aasi J.,  et~al., 2015, \mndoi [Classical and Quantum Gravity]
  {10.1088/0264-9381/32/11/115012}, 32, 115012

\bibitem[\protect\citeauthoryear{Abbott et~al.,}{Abbott
  et~al.}{2017a}]{PhysRevLett.119.161101}
Abbott B.~P.,  et~al., 2017a, \mndoi [Phys. Rev. Lett.]
  {10.1103/PhysRevLett.119.161101}, 119, 161101

\bibitem[\protect\citeauthoryear{Abbott et~al.,}{Abbott
  et~al.}{2017b}]{Abbott_2017a}
Abbott B.~P.,  et~al., 2017b, \mndoi [\apj] {10.3847/2041-8213/aa920c}, 848,
  L13

\bibitem[\protect\citeauthoryear{Abbott et~al.,}{Abbott
  et~al.}{2018}]{Abbott2018}
Abbott B.~P.,  et~al., 2018, \mndoi [Living Reviews in Relativity]
  {10.1007/s41114-018-0012-9}, 21, 3

\bibitem[\protect\citeauthoryear{Abbott et~al.,}{Abbott
  et~al.}{2019}]{Abbott_2018b}
Abbott B.~P.,  et~al., 2019, \mndoi [Phys. Rev. X] {10.1103/PhysRevX.9.031040},
  9, 031040

\bibitem[\protect\citeauthoryear{Abbott et~al.,}{Abbott
  et~al.}{2020}]{GW190814}
Abbott R.,  et~al., 2020, \mndoi [The Astrophysical Journal]
  {10.3847/2041-8213/ab960f}, 896, L44

\bibitem[\protect\citeauthoryear{Acernese et~al.,}{Acernese et~al.}{2014}]{AdV}
Acernese F.,  et~al., 2014, \mndoi [Classical and Quantum Gravity]
  {10.1088/0264-9381/32/2/024001}, 32, 024001

\bibitem[\protect\citeauthoryear{Adams et~al.,}{Adams et~al.}{2016}]{MBTA_2016}
Adams T.,  et~al., 2016, \mndoi [Classical and Quantum Gravity]
  {10.1088/0264-9381/33/17/175012}, 33, 175012

\bibitem[\protect\citeauthoryear{{Anderson} et~al.,}{{Anderson}
  et~al.}{2018}]{2018ApJ...864...22A}
{Anderson} M.~M.,  et~al., 2018, \mndoi [\apj] {10.3847/1538-4357/aad2d7},
  \href {https://ui.adsabs.harvard.edu/abs/2018ApJ...864...22A} {864, 22}

\bibitem[\protect\citeauthoryear{Aso, Michimura, Somiya, Ando, Miyakawa,
  Sekiguchi, Tatsumi  \& Yamamoto}{Aso et~al.}{2013}]{KAGRA_2}
Aso Y.,  Michimura Y.,  Somiya K.,  Ando M.,  Miyakawa O.,  Sekiguchi T.,
  Tatsumi D.,   Yamamoto H.,  2013, \mndoi [Phys. Rev. D]
  {10.1103/PhysRevD.88.043007}, 88, 043007

\bibitem[\protect\citeauthoryear{Bannister et~al.,}{Bannister
  et~al.}{2017}]{Bannister_2017}
Bannister K.~W.,  et~al., 2017, \mndoi [\apjl] {10.3847/2041-8213/aa71ff}, 841,
  L12

\bibitem[\protect\citeauthoryear{Bannister et~al.,}{Bannister
  et~al.}{2019}]{Bannister565}
Bannister K.~W.,  et~al., 2019, \mndoi [Science] {10.1126/science.aaw5903},
  365, 565

\bibitem[\protect\citeauthoryear{{Bochenek}, {Ravi}, {Belov}, {Hallinan},
  {Kocz}, {Kulkarni}  \& {McKenna}}{{Bochenek}
  et~al.}{2020}]{2020arXiv200510828B}
{Bochenek} C.~D.,  {Ravi} V.,  {Belov} K.~V.,  {Hallinan} G.,  {Kocz} J.,
  {Kulkarni} S.~R.,   {McKenna} D.~L.,  2020, \nat, \href
  {https://ui.adsabs.harvard.edu/abs/2020arXiv200510828B} {submitted,
  arXiv:2005.10828}

\bibitem[\protect\citeauthoryear{{Callister}, {Kanner}  \&
  {Weinstein}}{{Callister} et~al.}{2016}]{2016ApJ...825L..12C}
{Callister} T.,  {Kanner} J.,   {Weinstein} A.,  2016, \mndoi [\apjl]
  {10.3847/2041-8205/825/1/L12}, \href
  {https://ui.adsabs.harvard.edu/abs/2016ApJ...825L..12C} {825, L12}

\bibitem[\protect\citeauthoryear{Callister et~al.,}{Callister
  et~al.}{2019}]{Callister_2019}
Callister T.~A.,  et~al., 2019, \mndoi [The Astrophysical Journal]
  {10.3847/2041-8213/ab2248}, 877, L39

\bibitem[\protect\citeauthoryear{Cannon et~al.,}{Cannon
  et~al.}{2012}]{Cannon_2012}
Cannon K.,  et~al., 2012, \mndoi [The Astrophysical Journal]
  {10.1088/0004-637x/748/2/136}, 748, 136

\bibitem[\protect\citeauthoryear{{Cao}, {Yu}  \& {Zhou}}{{Cao}
  et~al.}{2018}]{2018ApJ...858...89C}
{Cao} X.-F.,  {Yu} Y.-W.,   {Zhou} X.,  2018, \mndoi [\apj]
  {10.3847/1538-4357/aabadd}, \href
  {https://ui.adsabs.harvard.edu/abs/2018ApJ...858...89C} {858, 89}

\bibitem[\protect\citeauthoryear{{Chatterjee}, {Ghosh}, {Brady}, {Kapadia},
  {Miller}, {Nissanke}  \& {Pannarale}}{{Chatterjee}
  et~al.}{2020}]{2020ApJ...896...54C}
{Chatterjee} D.,  {Ghosh} S.,  {Brady} P.~R.,  {Kapadia} S.~J.,  {Miller}
  A.~L.,  {Nissanke} S.,   {Pannarale} F.,  2020, \mndoi [\apj]
  {10.3847/1538-4357/ab8dbe}, \href
  {https://ui.adsabs.harvard.edu/abs/2020ApJ...896...54C} {896, 54}

\bibitem[\protect\citeauthoryear{{Chawla} et~al.,}{{Chawla}
  et~al.}{2017}]{2017ApJ...844..140C}
{Chawla} P.,  et~al., 2017, \mndoi [\apj] {10.3847/1538-4357/aa7d57}, \href
  {https://ui.adsabs.harvard.edu/abs/2017ApJ...844..140C} {844, 140}

\bibitem[\protect\citeauthoryear{{Chawla} et~al.,}{{Chawla}
  et~al.}{2020}]{2020arXiv200402862C}
{Chawla} P.,  et~al., 2020, arXiv e-prints, \href
  {https://ui.adsabs.harvard.edu/abs/2020arXiv200402862C} {in press,
  arXiv:2004.02862}

\bibitem[\protect\citeauthoryear{Chu, Howell, Rowlinson, Gao, Zhang, Tingay,
  Bo\"{e}r  \& Wen}{Chu et~al.}{2016}]{Chu_2016}
Chu Q.,  Howell E.~J.,  Rowlinson A.,  Gao H.,  Zhang B.,  Tingay S.~J.,
  Bo\"{e}r M.,   Wen L.,  2016, \mndoi [\mnras] {10.1093/mnras/stw576}, 459,
  121

\bibitem[\protect\citeauthoryear{{Cordes} \& {Chatterjee}}{{Cordes} \&
  {Chatterjee}}{2019}]{2019ARA&A..57..417C}
{Cordes} J.~M.,  {Chatterjee} S.,  2019, \mndoi [\araa]
  {10.1146/annurev-astro-091918-104501}, \href
  {https://ui.adsabs.harvard.edu/abs/2019ARA&A..57..417C} {57, 417}

\bibitem[\protect\citeauthoryear{{Dal Canton} \& {Harry}}{{Dal Canton} \&
  {Harry}}{2017}]{PyCBC_2017}
{Dal Canton} T.,  {Harry} I.~W.,  2017, arXiv e-prints, \href
  {https://ui.adsabs.harvard.edu/abs/2017arXiv170501845D} {p. arXiv:1705.01845}

\bibitem[\protect\citeauthoryear{Dobie, Murphy, Kaplan, Ghosh, Bannister  \&
  Hunstead}{Dobie et~al.}{2019a}]{dobie_2019}
Dobie D.,  Murphy T.,  Kaplan D.~L.,  Ghosh S.,  Bannister K.~W.,   Hunstead
  R.~W.,  2019a, \mndoi [Publications of the Astronomical Society of Australia]
  {10.1017/pasa.2019.9}, 36, e019

\bibitem[\protect\citeauthoryear{Dobie et~al.,}{Dobie
  et~al.}{2019b}]{Dobie_2019b}
Dobie D.,  et~al., 2019b, \mndoi [\apj] {10.3847/2041-8213/ab59db}, 887, L13

\bibitem[\protect\citeauthoryear{Fairhurst}{Fairhurst}{2009}]{Fairhurst_2009}
Fairhurst S.,  2009, \mndoi [New Journal of Physics]
  {10.1088/1367-2630/11/12/123006}, 11, 123006

\bibitem[\protect\citeauthoryear{Falcke \& Rezzolla}{Falcke \&
  Rezzolla}{2014}]{Falcke2014}
Falcke H.,  Rezzolla L.,  2014, \mndoi [A\&A] {10.1051/0004-6361/201321996},
  562, A137

\bibitem[\protect\citeauthoryear{Hansen \& Lyutikov}{Hansen \&
  Lyutikov}{2001}]{Hansen2001}
Hansen B. M.~S.,  Lyutikov M.,  2001, \mndoi [\mnras]
  {10.1046/j.1365-8711.2001.04103.x}, 322, 695

\bibitem[\protect\citeauthoryear{Hooper, Chung, Luan, Blair, Chen  \&
  Wen}{Hooper et~al.}{2012a}]{SPIIR1}
Hooper S.,  Chung S.~K.,  Luan J.,  Blair D.,  Chen Y.,   Wen L.,  2012a,
  \mndoi [Phys. Rev. D] {10.1103/PhysRevD.86.024012}, 86, 024012

\bibitem[\protect\citeauthoryear{Hooper et~al.,}{Hooper et~al.}{2012b}]{SPIIR2}
Hooper S.,  et~al., 2012b, \mndoi [Journal of Physics: Conference Series]
  {10.1088/1742-6596/363/1/012027}, 363, 012027

\bibitem[\protect\citeauthoryear{{James}, {Ekers}, {Macquart}, {Bannister}  \&
  {Shannon}}{{James} et~al.}{2019a}]{2019MNRAS.483.1342J}
{James} C.~W.,  {Ekers} R.~D.,  {Macquart} J.~P.,  {Bannister} K.~W.,
  {Shannon} R.~M.,  2019a, \mndoi [\mnras] {10.1093/mnras/sty3031}, \href
  {https://ui.adsabs.harvard.edu/abs/2019MNRAS.483.1342J} {483, 1342}

\bibitem[\protect\citeauthoryear{James, Anderson, Wen, Bosveld, Chu, Kovalam,
  Slaven-Blair  \& Williams}{James et~al.}{2019b}]{james_2019}
James C.~W.,  Anderson G.~E.,  Wen L.,  Bosveld J.,  Chu Q.,  Kovalam M.,
  Slaven-Blair T.~J.,   Williams A.,  2019b, \mndoi [\mnras]
  {10.1093/mnrasl/slz129}, 489, L75

\bibitem[\protect\citeauthoryear{Johnston et~al.,}{Johnston
  et~al.}{2008}]{ASKAP}
Johnston S.,  et~al., 2008, \mndoi [Experimental Astronomy]
  {10.1007/s10686-008-9124-7}, 22, 151

\bibitem[\protect\citeauthoryear{{Kaplan} et~al.,}{{Kaplan}
  et~al.}{2015}]{2015ApJ...814L..25K}
{Kaplan} D.~L.,  et~al., 2015, \mndoi [\apjl] {10.1088/2041-8205/814/2/L25},
  \href {https://ui.adsabs.harvard.edu/abs/2015ApJ...814L..25K} {814, L25}

\bibitem[\protect\citeauthoryear{Kaplan, Murphy, Rowlinson, Croft, Wayth  \&
  Trott}{Kaplan et~al.}{2016}]{kaplan_2016}
Kaplan D.~L.,  Murphy T.,  Rowlinson A.,  Croft S.~D.,  Wayth R.~B.,   Trott
  C.~M.,  2016, \mndoi [Publications of the Astronomical Society of Australia]
  {10.1017/pasa.2016.43}, 33, e050

\bibitem[\protect\citeauthoryear{{LIGO Scientific Collaboration and Virgo
  Collaboration} et~al.}{{LIGO Scientific Collaboration and Virgo
  Collaboration} et~al.}{2019}]{GCN25324}
{LIGO Scientific Collaboration and Virgo Collaboration} et~al., 2019, GCN,
  25324, 1

\bibitem[\protect\citeauthoryear{Lai}{Lai}{2012}]{Lai_2012}
Lai D.,  2012, \mndoi [\apj] {10.1088/2041-8205/757/1/l3}, 757, L3

\bibitem[\protect\citeauthoryear{Lorimer, Bailes, McLaughlin, Narkevic  \&
  Crawford}{Lorimer et~al.}{2007}]{Lorimer777}
Lorimer D.~R.,  Bailes M.,  McLaughlin M.~A.,  Narkevic D.~J.,   Crawford F.,
  2007, \mndoi [Science] {10.1126/science.1147532}, 318, 777

\bibitem[\protect\citeauthoryear{{Lu} \& {Piro}}{{Lu} \&
  {Piro}}{2019}]{2019ApJ...883...40L}
{Lu} W.,  {Piro} A.~L.,  2019, \mndoi [\apj] {10.3847/1538-4357/ab3796}, \href
  {https://ui.adsabs.harvard.edu/abs/2019ApJ...883...40L} {883, 40}

\bibitem[\protect\citeauthoryear{{Lu}, {Kumar}  \& {Narayan}}{{Lu}
  et~al.}{2019}]{2019MNRAS.483..359L}
{Lu} W.,  {Kumar} P.,   {Narayan} R.,  2019, \mndoi [\mnras]
  {10.1093/mnras/sty2829}, \href
  {https://ui.adsabs.harvard.edu/abs/2019MNRAS.483..359L} {483, 359}

\bibitem[\protect\citeauthoryear{Lyutikov}{Lyutikov}{2013}]{Lyutikov_2013}
Lyutikov M.,  2013, \mndoi [\apj] {10.1088/0004-637x/768/1/63}, 768, 63

\bibitem[\protect\citeauthoryear{{Macquart} \& {Ekers}}{{Macquart} \&
  {Ekers}}{2018}]{2018MNRAS.474.1900M}
{Macquart} J.~P.,  {Ekers} R.~D.,  2018, \mndoi [\mnras]
  {10.1093/mnras/stx2825}, \href
  {https://ui.adsabs.harvard.edu/abs/2018MNRAS.474.1900M} {474, 1900}

\bibitem[\protect\citeauthoryear{Macquart et~al.,}{Macquart
  et~al.}{2010}]{CRAFT}
Macquart J.-P.,  et~al., 2010, \mndoi [\pasa] {10.1071/AS09082}, 27, 272–282

\bibitem[\protect\citeauthoryear{{Margalit}, {Beniamini}, {Sridhar}  \&
  {Metzger}}{{Margalit} et~al.}{2020}]{2020arXiv200505283M}
{Margalit} B.,  {Beniamini} P.,  {Sridhar} N.,   {Metzger} B.~D.,  2020, \apjl,
  \href {https://ui.adsabs.harvard.edu/abs/2020arXiv200505283M} {submitted,
  arXiv:2005.05283}

\bibitem[\protect\citeauthoryear{{Mereghetti} et~al.,}{{Mereghetti}
  et~al.}{2020a}]{2020arXiv200506335M}
{Mereghetti} S.,  et~al., 2020a, arXiv e-prints, \href
  {https://ui.adsabs.harvard.edu/abs/2020arXiv200506335M} {p. arXiv:2005.06335}

\bibitem[\protect\citeauthoryear{{Mereghetti}, {Savchenko}, {Gotz}, {Ducci},
  {Ferrigno}, {Bozzo}, {Borkowski}  \& {Bazzano}}{{Mereghetti}
  et~al.}{2020b}]{2020GCN.27668....1M}
{Mereghetti} S.,  {Savchenko} V.,  {Gotz} D.,  {Ducci} L.,  {Ferrigno} C.,
  {Bozzo} E.,  {Borkowski} J.,   {Bazzano} A.,  2020b, GRB Coordinates Network,
  \href {https://ui.adsabs.harvard.edu/abs/2020GCN.27668....1M} {27668, 1}

\bibitem[\protect\citeauthoryear{Messick et~al.,}{Messick
  et~al.}{2017}]{GstLAL_2017a}
Messick C.,  et~al., 2017, \mndoi [Phys. Rev. D] {10.1103/PhysRevD.95.042001},
  95, 042001

\bibitem[\protect\citeauthoryear{Metzger \& Zivancev}{Metzger \&
  Zivancev}{2016}]{Metzger_2016}
Metzger B.~D.,  Zivancev C.,  2016, \mndoi [\mnras] {10.1093/mnras/stw1800},
  461, 4435

\bibitem[\protect\citeauthoryear{Mingarelli, Levin  \& Lazio}{Mingarelli
  et~al.}{2015}]{Mingarelli_2015}
Mingarelli C. M.~F.,  Levin J.,   Lazio T. J.~W.,  2015, \mndoi [The
  Astrophysical Journal] {10.1088/2041-8205/814/2/l20}, 814, L20

\bibitem[\protect\citeauthoryear{Nitz, Dal~Canton, Davis  \& Reyes}{Nitz
  et~al.}{2018}]{PyCBC_2018}
Nitz A.~H.,  Dal~Canton T.,  Davis D.,   Reyes S.,  2018, \mndoi [Phys. Rev. D]
  {10.1103/PhysRevD.98.024050}, 98, 024050

\bibitem[\protect\citeauthoryear{{Pilia} et~al.,}{{Pilia}
  et~al.}{2020}]{2020ApJ...896L..40P}
{Pilia} M.,  et~al., 2020, \mndoi [\apjl] {10.3847/2041-8213/ab96c0}, \href
  {https://ui.adsabs.harvard.edu/abs/2020ApJ...896L..40P} {896, L40}

\bibitem[\protect\citeauthoryear{Pshirkov \& Postnov}{Pshirkov \&
  Postnov}{2010}]{Pshirkov_2010}
Pshirkov M.,  Postnov K.,  2010, \apss, 330, 13

\bibitem[\protect\citeauthoryear{Ravi}{Ravi}{2019}]{Ravi2019}
Ravi V.,  2019, \mndoi [Nature Astronomy] {10.1038/s41550-019-0831-y}, 3, 928

\bibitem[\protect\citeauthoryear{Ravi \& Lasky}{Ravi \&
  Lasky}{2014}]{Ravi_2014}
Ravi V.,  Lasky P.~D.,  2014, \mndoi [\mnras] {10.1093/mnras/stu720}, 441, 2433

\bibitem[\protect\citeauthoryear{{Ridnaia} et~al.,}{{Ridnaia}
  et~al.}{2020a}]{2020arXiv200511178R}
{Ridnaia} A.,  et~al., 2020a, arXiv e-prints, \href
  {https://ui.adsabs.harvard.edu/abs/2020arXiv200511178R} {p. arXiv:2005.11178}

\bibitem[\protect\citeauthoryear{{Ridnaia} et~al.,}{{Ridnaia}
  et~al.}{2020b}]{2020GCN.27669....1R}
{Ridnaia} A.,  et~al., 2020b, GRB Coordinates Network, \href
  {https://ui.adsabs.harvard.edu/abs/2020GCN.27669....1R} {27669, 1}

\bibitem[\protect\citeauthoryear{{Rowlinson} \& {Anderson}}{{Rowlinson} \&
  {Anderson}}{2019}]{2019MNRAS.489.3316R}
{Rowlinson} A.,  {Anderson} G.~E.,  2019, \mndoi [\mnras]
  {10.1093/mnras/stz2295}, \href
  {https://ui.adsabs.harvard.edu/abs/2019MNRAS.489.3316R} {489, 3316}

\bibitem[\protect\citeauthoryear{{Rowlinson} et~al.,}{{Rowlinson}
  et~al.}{2016}]{2016MNRAS.458.3506R}
{Rowlinson} A.,  et~al., 2016, \mndoi [\mnras] {10.1093/mnras/stw451}, \href
  {https://ui.adsabs.harvard.edu/abs/2016MNRAS.458.3506R} {458, 3506}

\bibitem[\protect\citeauthoryear{{Sachdev} et~al.,}{{Sachdev}
  et~al.}{2019}]{GstLAL_2019b}
{Sachdev} S.,  et~al., 2019, \prd, \href
  {https://ui.adsabs.harvard.edu/abs/2019arXiv190108580S} {submitted,
  arXiv:1901.08580}

\bibitem[\protect\citeauthoryear{{Scholz} \& {Chime/Frb
  Collaboration}}{{Scholz} \& {Chime/Frb
  Collaboration}}{2020}]{2020ATel13681....1S}
{Scholz} P.,  {Chime/Frb Collaboration} 2020, The Astronomer's Telegram, \href
  {https://ui.adsabs.harvard.edu/abs/2020ATel13681....1S} {13681, 1}

\bibitem[\protect\citeauthoryear{Shannon et~al.,}{Shannon
  et~al.}{2018}]{Shannon2018}
Shannon R.~M.,  et~al., 2018, \mndoi [Nature] {10.1038/s41586-018-0588-y}, 562,
  386

\bibitem[\protect\citeauthoryear{Singer \& Price}{Singer \&
  Price}{2016}]{Singer_bayestar}
Singer L.~P.,  Price L.~R.,  2016, \mndoi [Phys. Rev. D]
  {10.1103/PhysRevD.93.024013}, 93, 024013

\bibitem[\protect\citeauthoryear{Singer et~al.,}{Singer
  et~al.}{2014}]{Singer_2014}
Singer L.~P.,  et~al., 2014, \mndoi [The Astrophysical Journal]
  {10.1088/0004-637x/795/2/105}, 795, 105

\bibitem[\protect\citeauthoryear{{Sokolowski} et~al.,}{{Sokolowski}
  et~al.}{2018}]{2018ApJ...867L..12S}
{Sokolowski} M.,  et~al., 2018, \mndoi [\apjl] {10.3847/2041-8213/aae58d},
  \href {https://ui.adsabs.harvard.edu/abs/2018ApJ...867L..12S} {867, L12}

\bibitem[\protect\citeauthoryear{Somiya}{Somiya}{2012}]{KAGRA_1}
Somiya K.,  2012, \mndoi [Classical and Quantum Gravity]
  {10.1088/0264-9381/29/12/124007}, 29, 124007

\bibitem[\protect\citeauthoryear{Spitler et~al.,}{Spitler
  et~al.}{2016}]{FRB121102}
Spitler L.~G.,  et~al., 2016, \mndoi [Nature] {10.1038/nature17168}, 531, 202

\bibitem[\protect\citeauthoryear{Thornton et~al.,}{Thornton
  et~al.}{2013}]{Thornton53}
Thornton D.,  et~al., 2013, \mndoi [Science] {10.1126/science.1236789}, 341, 53

\bibitem[\protect\citeauthoryear{Tingay et~al.,}{Tingay et~al.}{2013}]{MWA}
Tingay S.~J.,  et~al., 2013, \mndoi [\pasa] {10.1017/pasa.2012.007}, 30, e007

\bibitem[\protect\citeauthoryear{{Tingay} et~al.,}{{Tingay}
  et~al.}{2015}]{2015AJ....150..199T}
{Tingay} S.~J.,  et~al., 2015, \mndoi [\aj] {10.1088/0004-6256/150/6/199},
  \href {https://ui.adsabs.harvard.edu/abs/2015AJ....150..199T} {150, 199}

\bibitem[\protect\citeauthoryear{Totani}{Totani}{2013}]{Totani_2013}
Totani T.,  2013, \mndoi [\pasj] {10.1093/pasj/65.5.L12}, 65

\bibitem[\protect\citeauthoryear{Wang, Yang, Wu, Dai  \& Wang}{Wang
  et~al.}{2016}]{Wang_2016}
Wang J.-S.,  Yang Y.-P.,  Wu X.-F.,  Dai Z.-G.,   Wang F.-Y.,  2016, \mndoi
  [\apjl] {10.3847/2041-8205/822/1/l7}, 822, L7

\bibitem[\protect\citeauthoryear{Wang, Peng, Wu  \& Dai}{Wang
  et~al.}{2018}]{Wang_2018}
Wang J.-S.,  Peng F.-K.,  Wu K.,   Dai Z.-G.,  2018, \mndoi [\apj]
  {10.3847/1538-4357/aae531}, 868, 19

\bibitem[\protect\citeauthoryear{{Zhang} et~al.,}{{Zhang}
  et~al.}{2020}]{2020ATel13687....1Z}
{Zhang} S.~N.,  et~al., 2020, The Astronomer's Telegram, \href
  {https://ui.adsabs.harvard.edu/abs/2020ATel13687....1Z} {13687, 1}

\makeatother
\end{thebibliography}

\end{document}